\documentstyle[aps,prl,preprint,floats,epsfig]{revtex}

\begin{document}

\preprint{\tighten\vbox{\hbox{\hfil CLNS 01/1736}
                        \hbox{\hfil CLEO 01-10}
}}

\title{Search for the Familon via\\ $B^{\pm} \to \pi^{\pm}X^0$,
$B^{\pm} \to K^{\pm}X^0$, and $B^0 \to K^0_S X^0$ Decays}  

\author{CLEO Collaboration}
\date{June 6, 2001}

\maketitle
\tighten

\begin{abstract} 
We have searched for the two-body decay 
of the $B$ meson to a light pseudoscalar meson 
$h=\pi^{\pm},K^{\pm},K^0_S$ and a massless neutral 
weakly-interacting particle $X^0$ such as the familon, 
the Nambu-Goldstone boson associated with a spontaneously broken 
global family symmetry. We find no significant signal by analyzing 
a data sample containing 9.7 million $B\bar{B}$ mesons collected
with the CLEO detector at the Cornell Electron Storage Ring, and
set a 90\% C.L. upper limit of $4.9 \times 10^{-5}$ and
$5.3 \times 10^{-5}$ on the branching fraction for the decays
$B^{\pm} \to h^{\pm}X^0$ and $B^{0}\to K_{S}^{0}X^0$, respectively. 
These upper limits correspond to a lower bound of $\approx 10^{8}$ GeV 
on the family symmetry breaking scale involving the third 
generation of quarks.
\end{abstract}
\newpage

{
\renewcommand{\thefootnote}{\fnsymbol{footnote}}

\begin{center}
R.~Ammar,$^{1}$ A.~Bean,$^{1}$ D.~Besson,$^{1}$ X.~Zhao,$^{1}$
S.~Anderson,$^{2}$ V.~V.~Frolov,$^{2}$ Y.~Kubota,$^{2}$
S.~J.~Lee,$^{2}$ R.~Poling,$^{2}$ A.~Smith,$^{2}$
% ...
\end{center}
 
\small
\begin{center}
$^{1}${University of Kansas, Lawrence, Kansas 66045}\\
$^{2}${University of Minnesota, Minneapolis, Minnesota 55455}\\
% ...
\end{center}
 
\setcounter{footnote}{0}
}
\newpage
The origin of family replication remains one of the major
puzzles in particle physics. Why do we have three families 
of fermions, which are indistinguishable with respect 
to the strong and electroweak interactions? Neither 
the Standard Model (even incorporating the Higgs mechanism)
nor its extension by various unification schemes in the 
framework of one family (SU(5), SO(10)) is able to provide 
a deep physical reason for the existence of the mass 
hierarchy among the generations and the weak mixing of 
quarks and leptons. In the absence of a concrete model, 
it is natural to assume 
that the underlying theory possesses a ``horizontal'' family 
symmetry which is spontaneously broken at some large 
energy scale. Among several possibilities, the most 
attractive is the assumption of a global (and continuous) 
flavor symmetry \cite{theory1}. This symmetry, under some
conditions \cite{chang}, automatically induces the 
Peccei-Quinn symmetry \cite{PQsym}, and thus provides a 
solution for the strong $CP$ problem. The spontaneous 
symmetry breaking of a continuous and global family 
symmetry implies the existence of neutral massless 
Nambu-Goldstone bosons \cite{Goldstone}, called familons, 
which can have flavor-conserving as well as 
flavor-changing couplings with the fermions \cite{theory1,familon}.

Flavor-changing couplings between the familon and fermions 
induce decays, such as $K^{+}\to\pi^{+}X^0$ or 
$\mu^{+}\to e^{+}(\gamma)X^0$, that have been studied 
experimentally \cite{k+pi+f,mu+e+f}. Upper limits on the rate 
of these decays led to the lower bounds on the family symmetry 
breaking scale involving the first two generations: 
$\approx$ $10^{11}$ GeV and $\approx$ $10^{9}$ GeV in the 
hadronic and the leptonic sector, respectively. In contrast, 
bounds on the flavor scale involving the third generation 
are less thoroughly studied experimentally,
although, some theoretical models suggest that the familon 
couples preferentially to the third generation \cite{theory2}.
The upper limits for $\tau \to \ell X^0$ ($\ell = e, \mu$) 
\cite{taumuf} led to a lower bound on the family symmetry 
breaking scale $F \gtrsim 10^{6}$ GeV in the leptonic 
sector, and no bounds have been reported in the hadronic sector.

Familon couplings to the third generation are also of 
interest from a cosmological point of view. A massive 
unstable neutrino (typically the tau-neutrino) was proposed 
to decay into a lighter neutrino and a massless boson, 
such as a familon, in several cosmological scenarios 
related to big-bang nucleosynthesis \cite{bbn}, and large 
scale structure formation \cite{cosmology}, in order to 
obtain a reasonable agreement between theory and observation. 
Since the process $\nu_{\tau} \to \nu_{l}f$ is 
related to the decay modes $\tau\to \ell f$ and 
$b\to q_d f$ ($q_d = d,s$) through SU(2)$_L$ and SU(5) GUT 
gauge symmetries, searches for the latter decay modes can 
test the cosmological scenarios as well \cite{familon}.

The decay of the $b$ quark $b\to q_d f$ would 
lead to the decay $B\to hf$ ($h=\pi,K$) through vector 
coupling and $B\to Vf$ ($V=\rho,K^*$) through 
axial coupling, respectively. The purpose of this study 
is to search for the $B^{\pm} \to h^{\pm}X^0$ 
and $B^{0} \to K_{S}^{0} X^0$ decays, where $X^0$ is 
any neutral massless weakly-interacting particle including 
the familon, using the CLEO data set. The lack of a signal
allows us to obtain a constraint on the vector coupling of 
the familon to third generation hadrons for the first time. 
(The analysis is sensitive to new physics including massless 
weakly-interacting neutral particles as well.) The partial width 
$\Gamma$ of the decay is related to the family symmetry breaking 
scale $F$ through the formula
\begin{equation}
\Gamma(B \to hf)=\frac{M_{B}^{3}}{16\pi}\left(
1-\frac{m_{h}^{2}}{M_{B}^{2}}\right)
^{3}\frac{g_{V}^{2}T_{bd(s)}^{2}}{F^{2}}\mid F_{1}(0)\mid^{2},
\label{Eq:Gamma}
\end{equation}
where $M_B$, $m_{h}$ are the masses of the mesons 
involved in the decay process, $g_V$ is the vector type 
coupling constant, $T_{bd(s)}$ are the generators of the 
broken symmetry, and $F_{1}(0)$ is the form factor \cite{familon}.

The data analyzed in this study were collected with the 
CLEO detector at the Cornell Electron Storage Ring (CESR), 
a symmetric $e^+ e^-$ collider. The components of the 
detector \cite{cleo2} most relevant to this analysis are the 
charged particle tracking system, the CsI electromagnetic 
calorimeter, and the muon detector. Trajectories of charged 
particles were reconstructed using a system of three concentric 
wire chambers (a 6-layer straw tube chamber, a 10-layer 
precision drift chamber, and a 51-layer main drift chamber) 
covering $95\%$ of the total solid angle, operating 
in an axial solenoidal magnetic field of $1.5$ T. The main 
drift chamber also provided a measurement of the specific
ionization loss ($dE/dx$) used for particle identification.
Photons were detected by a CsI(Tl) electromagnetic 
calorimeter covering $98\%$ of $4\pi$. The muon chambers 
consisted of proportional counters embedded at various
depths in the steel absorber. Approximately 2/3 of the data 
were collected with an upgraded detector, 
in which the innermost straw tube chamber was replaced 
with a three-layer, double-sided silicon vertex detector
\cite{cleo25}, and the gas in the main drift chamber was 
changed from an argon-ethane to a helium-propane mixture. 
These modifications led to an improved particle 
identification and momentum resolution.

The results in this Letter are based upon an integrated 
luminosity of $9.2$ fb$^{-1}$ of $\ e^{+}e^{-}$ data
corresponding to 9.7 million $B\bar{B}$ meson pairs
collected at the $\Upsilon$(4S) resonance energy of 
$10.58$ GeV (``on-resonance sample'') and $4.6$ fb$^{-1}$ 
at $60$ MeV below the $\Upsilon$(4S) resonance
(``off-resonance sample''). The study of the off-resonance 
sample enables us to statistically subtract the continuum
($e^+e^- \to q\bar{q}$, $q = u, d, c, s$)
background contribution from the on-resonance sample.
In order to study signal reconstruction efficiency and to
optimize selection criteria we generated Monte Carlo
simulated samples with a GEANT-based \cite{geant} 
simulation of the CLEO detector response. Simulated 
data samples were processed in a similar manner as the data.

The experimental signature of the the 
$B^{\pm} \to h^{\pm}X^0$ and
$B^{0} \to K_{S}^{0}X^0$ decays is that 
the familon as a neutral very weakly-interacting 
particle escapes from the detector without any trace
and only its light meson partner can be observed. 
Due to the two-body decay structure, the meson 
partner is produced with a well defined momentum 
of $2.65$ GeV/$c$ in the center of mass frame 
of the decaying $B$ meson. However, in the lab 
frame its momentum is spread between 
$2.49-2.80$ GeV/$c$ due to Doppler broadening.
Other detected particles and photons must be 
coming from the decay of the other $B$ meson. 
Our analysis strategy to search for these decay 
modes is the following: (1) we select events with 
a well identified light meson having a momentum 
in the expected range while (2) all remaining
particles must be consistent with the decay of a 
second $B$ meson, and (3) eliminate as much 
continuum background as possible.

Candidates for the 
$\pi^{\pm}$ or $K^{\pm}$ meson partner of the 
familon (``meson candidate'') were selected from 
well-reconstructed tracks originating near the 
$e^+e^-$ interaction point (IP). Since charged 
$\pi$ and $K$ meson separation in the momentum 
range expected is difficult with the 
CLEO detector, we combined the charged 
$B$ decay modes by requiring the charged meson 
candidate's $dE/dx$ to be consistent with either 
the pion or the kaon hypothesis within $2.5$ standard 
deviation ($\sigma$). We rejected electrons based on 
$dE/dx$ and the ratio of the track momentum to the 
associated shower energy deposited in the CsI calorimeter.
Muons were rejected based on the penetration depth in 
the steel absorber surrounding the detector. 
The $K_S^0$ candidates were reconstructed via their 
decay into $\pi^+\pi^-$ by requiring a decay vertex 
displacement from the IP and an invariant $\pi\pi$ 
mass within $10$ MeV/$c^2$ of the known $K_S^0$ mass.
We accepted meson candidates with momentum in the 
range $2.49<p_{h^{\pm}}<2.81$ GeV/$c$ 
or $2.47<p_{K^0_S}<2.79$ GeV/$c$.
These and other selection criteria were optimized 
by maximizing the signal significance, $S^2/(S+B)$, 
where $S$ and $B$, the expected signal and background 
level was determined from Monte-Carlo simulated samples 
assuming a signal branching fraction of $10^{-5}$.

Since the remaining particles in the event must 
originate from the decay of the second $B$ meson 
we required that the beam constrained mass, 
$M(B) = \sqrt{E_{\rm beam}^2 - (\sum {\bf p}_i)^2)}$, be 
close to the $B$ meson mass and the energy difference, 
$\Delta E = \sum E_i - E_{\rm beam}$, be close to zero, where 
$E_i$ and ${\bf p}_i$ are the energy and momentum
of all detected particles in the event except for
the meson candidate.
The optimization of the selection 
criteria on the $M(B)$ and $\Delta E$ variables resulted 
in $M(B)>5.245$ GeV/$c^2$ ($M(B)>5.24$ GeV/$c^2$) 
and $-2.1<\Delta E<0.3$ GeV ($-3.0<\Delta E<0.4$ GeV)
limits for the charged (neutral) $B$ decay mode.

The main contribution to the background comes from continuum 
events. These events typically exhibit a 
two-jet structure and produce high momentum back-to-back
tracks, while $B\bar{B}$ events tend to have a more isotropic 
decay structure, since the $B$ mesons are produced nearly at rest
($P_B \approx 0.32$ GeV/$c$). We used the Fisher discriminant 
technique \cite{fisher} to reduce the continuum background. 
The Fisher discriminant was formed as the linear combination 
of 14 shape variables: 9 momentum flow variables (the sum of 
the momentum of all detected particles in $10^o$ angular bins 
around the direction of the meson candidate); the angle between 
the momentum of the other $B$ meson reconstructed from the 
rest of the event and the $e^+e^-$ collision (``beam'') 
axis; the angle between the momentum of the meson candidate 
and the beam axis; the second order normalized Fox-Wolfram 
moment \cite{r2}; the angle between the momentum of the meson 
candidate and the thrust axis of the rest of the event;
and the maximum opening angle of the cone opposite to the 
momentum of the meson candidate, in which no other charged 
track, $\pi^0$ or $K_S^0$ was detected.
The combination coefficients were chosen to maximize the 
separation between the simulated signal and continuum 
background samples.

\begin{figure}[t]
\begin{center}
\includegraphics[height=3.0in]{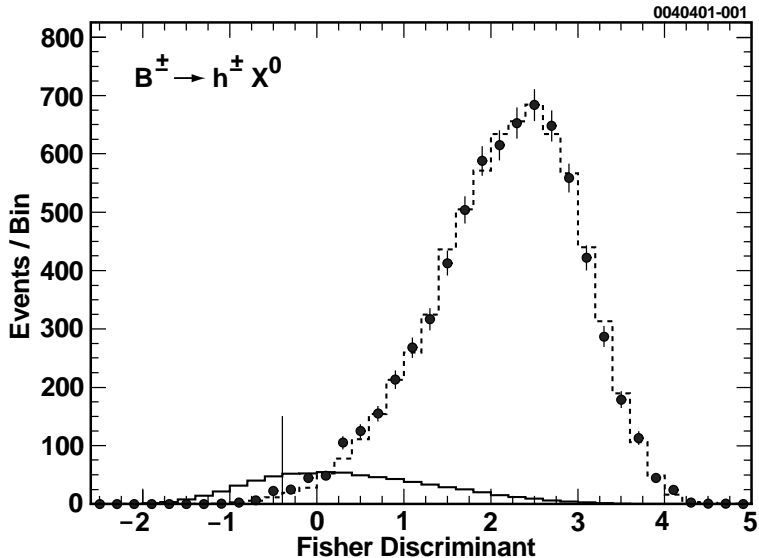}
\caption{Distribution of the Fisher discriminant used in the 
$B^{\pm}\to h^{\pm}X^0$ analysis for simulated 
signal (solid) and continuum (dashed) as well as off-resonance data
(points) samples. The histograms are normalized to the statistics of 
the off-resonance data. The signal histogram is plotted assuming a
branching ratio of $50 \times 10^{-5}$. The vertical line represents 
the optimum selection value below which events were accepted.}
\label{fisher}
\end{center}
\end{figure}

The distribution of the Fisher discriminant used in 
the charged $B$ analysis is shown for simulated events
and off-resonance data on Fig.\ \ref{fisher}. 
The agreement between simulated continuum and off-resonance 
events is very good. We selected candidate events with a 
Fisher discriminant less than $-0.4$ in case of the charged 
$B$ decay mode, and less than $1.0$ for the neutral $B$ 
decay mode.

The overall signal selection efficiency is
$7.2\%$ for $B^{\pm} \to h^{\pm}X^0$ 
and $6.6\%$ for the $B^{0} \to K_{S}^{0}X^0$ events. 
The systematic error on the efficiency is 
$13\%$ ($18\%$) for the charged (neutral) $B$ decay mode. 
The contributions to this error are due to the
uncertainties in the tracking efficiency, $2\%$ ($4\%$), 
the momentum selection, $1\%$ ($1\%$), $M(B)$ and 
$\Delta E$ selection, $6\%$ ($6\%$), Fisher discriminant
restriction, $11\%$ ($16\%$), and limited
Monte Carlo statistics, $1\%$ ($1\%$).

\begin{figure}[t]
\begin{center}
\includegraphics[height=4.5in]{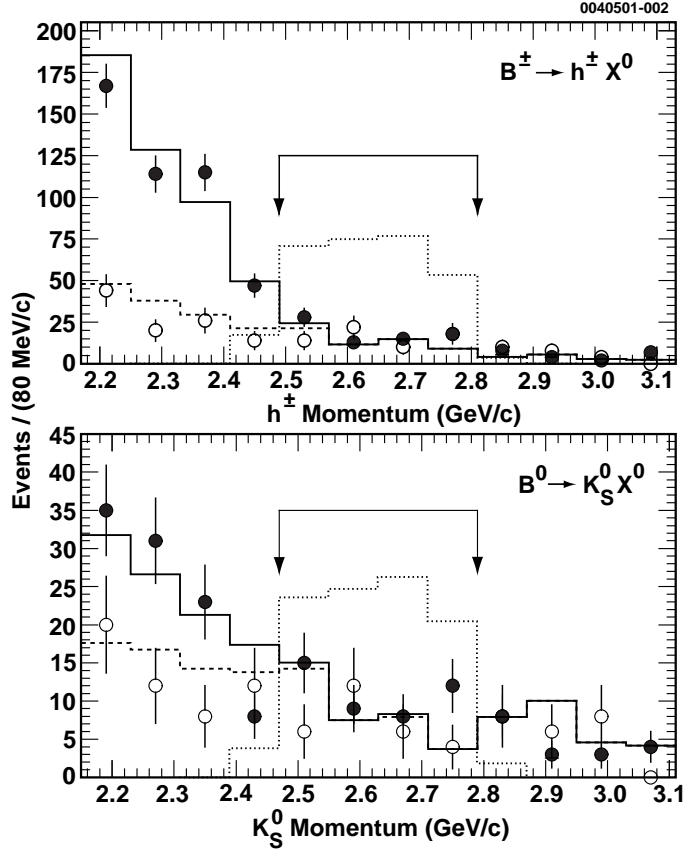}
\caption{Momentum distribution of the meson candidates. Filled and empty 
dots represent the on-resonance and the normalized off-resonance data, 
respectively. Solid histogram shows the prediction from 
$e^+ e^- \to q{\bar q}$ plus $b\to c$ simulations while 
the dashed histogram shows the distribution from 
$e^+ e^- \to q{\bar q}$ only. 
These histograms are normalized to the statistics of our data sample. 
Simulated signal events are shown by the dotted histogram assuming that 
${\cal B}(B^{\pm}\to h^{\pm}X^0)\approx 30\times 10^{-5}$
and ${\cal B}(B^{0}\to K_{S}^{0}X^0)\approx 12\times 10^{-5}$. 
The accepted signal region is indicated by the arrows.}
\label{result}
\end{center}
\end{figure}

Table~\ref{table} lists the number of events passing 
the consecutive selection requirements in the data and 
simulated signal samples. Figure~\ref{result} shows the 
momentum distribution of the meson candidate for 
on-resonance and off-resonance events along with the 
distributions for simulated events after all selection 
criteria except the tight momentum restriction on the meson 
candidate were applied. The number of on-resonance (off-resonance) 
events in the selected momentum range is 74 (32)
in case of the $B^{\pm} \to h^{\pm}X^0$ 
and 44 (14) in case of the $B^{0} \to K_{S}^{0}X^0$ 
analysis. The study of the background from $b\to c$, 
and other rare $b\to u$ and $b\to s$
decays as well as from tau decays using simulated data 
samples showed these to be negligible.

\begin{table}[b]
\caption{Number of events passing each consecutive selection criteria 
in on-resonance, off-resonance data and simulated signal samples.}
\label{table}
\begin{tabular}{ccccccc}
\multicolumn{1}{c}{}&
\multicolumn{3}{c}{$B^{\pm} \to h^{\pm}X^0$}&
\multicolumn{3}{c}{$B^0 \to K^0_S X^0$}\\
\cline{2-4}\cline{5-7}
&On-res.&Off-res.&MC Signal&On-res.&Off-res.& MC Signal\\
\tableline
Total events&57 million&23 million&180,000&57 million&23 million& 90,000\\
Pre-selected events&157,919&73,671&90,211&64,207&31,230&36,953\\
Momentum selection&41,981&20,437&83,592&18,675&9,224&34,720\\
$M_B$ and $\Delta E$ selection&14,243&7,073&55,024&2,330&1,135&17,725\\
Fisher selection&74&32&12,896&44&14&5,973\\
\end{tabular}
\end{table}

We calculated the branching fraction based on
\begin{equation}
{\cal B} = \frac{N_{\rm on} - RN_{\rm off}}{\epsilon N_B},
\end{equation}
where $N_{\rm on}$ and $N_{\rm off}$ are the observed 
events in the signal region in the on-resonance and 
off-resonance data samples, respectively, $R$($=2.0$) 
is the normalization coefficient between the two 
samples, $\epsilon$ is the signal selection efficiency, and 
$N_{B}$ is the total number of charged (neutral) $B$ mesons 
in the data sample, assuming equal production of charged 
and neutral $B$ meson pairs from the 
$\Upsilon (4S)$ \cite{4SBB}. We find 
${\cal B}(B^{\pm}\to h^{\pm}X^0)=(1.4\pm 2.1)\times 10^{-5}$
and
${\cal B}(B^{0}\to K_{S}^{0}X^0)=(2.5\pm 1.7)\times 10^{-5}$.
The error in the branching fraction is dominated by the statistical 
error in $N_{\rm on}$ and $N_{\rm off}$.
We derived a $90\%$ confidence level upper limit based on the 
frequentist approach applied for Gaussian data close to a physical 
boundary \cite{UL}: 
${\cal B}(B^{\pm}\to h^{\pm}X^0)<4.9\times10^{-5}$
and ${\cal B}(B^{0}\to K_{S}^{0}X^0)<5.3\times10^{-5}$.

The upper limits can be converted into a lower bound on 
the family symmetry breaking scale, $F^V_{bs(d)}=F/(g^{V}T_{bs(d)})$, 
with vector-like coupling between the familon and 
the quarks using Eq.\ \ref{Eq:Gamma}. To do so we take 
the form factor $F_1(0)$ to be $0.25$ from a sum rules
calculation \cite{form_factor}. The upper limit on the
branching fraction of $B^0 \to K^0_S X^0$ 
gives $F^V_{bs}\gtrsim 6.4\times 10^7$ GeV. 
The other limit gives a slightly better bound of 
$F^V_{bs(d)}\gtrsim 1.3\times 10^8$ GeV with the 
assumption that the familon couples to the $d$ and $s$ 
quark with approximately the same strength 
($F_{bs}\approx F_{bd}$).

In conclusion, we performed a search for the decays
$B^{\pm}\to h^{\pm}X^0$ and $B^{0}\to
K_{S}^{0}X^0$, setting upper limits for the branching
fractions at $4.9 \times 10^{-5}$ and $5.3 \times 10^{-5}$
respectively. These limits constrain new physics leading to
two-body $B$ decays involving any massless neutral 
weakly-interacting particle $X^0$. Applying the limit to the
case where $X^0$ is a familon, we obtain the first lower bound
on the family symmetry breaking scale involving the third 
generation of quarks at $10^8$ GeV.

We gratefully acknowledge the effort of the CESR staff in providing us with
excellent luminosity and running conditions.
This work was supported by 
the National Science Foundation,
the U.S. Department of Energy,
the Research Corporation,
the Natural Sciences and Engineering Research Council of Canada
and the Texas Advanced Research Program.

\end{document}